\documentclass[aps,prb,twocolumn,superscriptaddress]{revtex4-2}
\usepackage{graphicx}
\usepackage{mathrsfs}
\usepackage[colorlinks=true,allcolors=blue]{hyperref}
\usepackage{bm}
\usepackage{amsmath}
\usepackage{dcolumn}
\usepackage{epstopdf}
\usepackage{dsfont}
\usepackage{amssymb}
\usepackage{tabularx}
\usepackage{array}
\usepackage{colordvi}
\usepackage{braket}
\usepackage{float}
\usepackage{mathtools}
\usepackage{upgreek}
\usepackage{booktabs}
\usepackage{multirow}

\begin{document}
\title{Giant electrode effect on tunneling magnetoresistance and electroresistance in van der Waals intrinsic multiferroic tunnel junctions using VS$_2$}
\author{Zhi Yan$^\ddagger$}
\email[Corresponding author:~]{yanzhi@sxnu.edu.cn}
\affiliation{School of Chemistry and Materials Science $\&$ Key Laboratory of Magnetic Molecules and Magnetic Information Materials of Ministry of Education, Shanxi Normal University, Taiyuan 030031, China}
\affiliation{Research Institute of Materials Science $\&$ Collaborative Innovation Center for Shanxi Advanced Permanent Magnetic Materials and Technology, Shanxi Normal University, Taiyuan 030031, China}
\thanks{These authors contributed equally to this work.}
\author{Ruixia Yang$^\ddagger$}
\affiliation{School of Chemistry and Materials Science $\&$ Key Laboratory of Magnetic Molecules and Magnetic Information Materials of Ministry of Education, Shanxi Normal University, Taiyuan 030031, China}
\thanks{These authors contributed equally to this work.}
\author{Cheng Fang}
\affiliation{School of Chemistry and Materials Science $\&$ Key Laboratory of Magnetic Molecules and Magnetic Information Materials of Ministry of Education, Shanxi Normal University, Taiyuan 030031, China}
\author{Wentian Lu}
\affiliation{School of Chemistry and Materials Science $\&$ Key Laboratory of Magnetic Molecules and Magnetic Information Materials of Ministry of Education, Shanxi Normal University, Taiyuan 030031, China}
\author{Xiaohong Xu}
\email[Corresponding author:~]{xuxh@sxnu.edu.cn}
\affiliation{School of Chemistry and Materials Science $\&$ Key Laboratory of Magnetic Molecules and Magnetic Information Materials of Ministry of Education, Shanxi Normal University, Taiyuan 030031, China}
\affiliation{Research Institute of Materials Science $\&$ Collaborative Innovation Center for Shanxi Advanced Permanent Magnetic Materials and Technology, Shanxi Normal University, Taiyuan 030031, China}

\date{\today{}}

\begin{abstract}
Van der Waals multiferroic tunnel junctions (vdW-MFTJs) with multiple nonvolatile resistive states are highly suitable for new physics and next-generation storage electronics.
However, currently reported vdW-MFTJs are based on two types of materials, i.e., vdW ferromagnetic and ferroelectric materials, forming a multiferroic system. This undoubtedly introduces additional interfaces, increasing the complexity of experimental preparation. Herein, we engineer vdW intrinsic MFTJs utilizing bilayer VS$_2$. By employing the nonequilibrium Green’s function combined with density functional theory, we systematically investigate the influence of three types of electrodes (including non-vdW pure metal Ag/Au, vdW metallic 1T-MoS$_2$/2H-PtTe$_2$, and vdW ferromagnetic metallic Fe$_3$GaTe$_2$/Fe$_3$GeTe$_2$) on the electronic transport properties of VS$_2$-based intrinsic MFTJs. We demonstrate that these MFTJs manifest a giant electrode-dependent electronic transport characteristic effect. 
Comprehensively comparing these electrode pairs, the Fe$_3$GaTe$_2$/Fe$_3$GeTe$_2$ electrode combination exhibits optimal transport properties, the maximum TMR (TER) can reach 10949\% (69\%) and the minimum resistance-area product (RA) is 0.45 $\Omega$·$\mu$m$^{2}$, as well as the perfect spin filtering and negative differential resistance effects.
More intriguingly, TMR (TER) can be further enhanced to 34000\% (380\%) by applying an external bias voltage, while RA can be reduced to 0.16 $\Omega$·$\mu$m$^{2}$ under the influence of biaxial stress.
Additionally, considering the impact of surface dangling bonds of pure metal electrodes on the multiferroicity of VS$_2$, we introduce a graphene interlayer between them. This strategy effectively preserves the intrinsic properties of VS$_2$ and significantly amplifies the TMR (TER) of the MFTJ composed of Ag/Au electrode pairs by an order of magnitude.
Our proposed concept of designing vdW-MFTJs using intrinsic multiferroic materials points towards new avenues in experimental exploration.

\end{abstract}

\maketitle
\section{Introduction}
The combination of ferromagnetism, ferroelectricity, or multiferroicity with quantum tunneling effects can give rise to novel spintronics devices, such as magnetic, ferroelectric, and multiferroic tunnel junctions (MFTJs)~\cite{miyazaki1995giant, Kohlstedt2005FTJ, Duan2006MFTJ}. These devices offer significant advantages in terms of energy efficiency, performance, and data storage capabilities.
However, traditional tunnel junction devices fabricated from three-dimensional perovskite-oxide materials are often constrained by quantum size effects and typically exhibit large resistance-area product (RA)~\cite{borisov2015spin,velev2009magnetic}, limiting their ability to meet the demands for higher storage density, faster read/write speeds, and lower power consumption in miniaturized multifunctional electronic devices.
In recent years, the emergence of two-dimensional van der Waals ferromagnetic~\cite{deng2018gate,zhang2022above,huang2017layer} and ferroelectric materials~\cite{Ding2017NC,liu2016room} has paved the way for the experimental preparation of multifunctional miniaturized tunnel junction devices~\cite{MTJ0, MTJ1, MTJ2, MTJ3, MTJ4, MTJ5, MTJ7, MTJ8, Li2021PRA, FTJ0, FTJ2, FTJ3}. Among these, van der Waals multiferroic tunnel junctions (vdW-MFTJs) stand out prominently due to their combined tunneling magnetoresistance (TMR) and tunneling electroresistance (TER) effects, enabling the realization of multiple non-volatile resistive states.

The concept of vdW-MFTJs was formally introduced by Su et al.~\cite{Su2020NL}, based on the Fe$_n$GeTe$_2$/In$_2$Se$_3$($n$ = 3, 4, 5) heterostructure. Subsequently, Hu et al.~\cite{hu2021first} proposed MX$_2$(M=Mn, V, Cr; X=Se, Te)/In$_2$Se$_3$-based MFTJs. Following that, Bai et al.~\cite{bai2021van} reported low RA in CrSe$_2$/CuInP$_2$S$_6$/CrSe$_2$ vdW-MFTJ. Then we also designed vdW-MFTJs with six non-volatile resistive states based on Fe$_3$GeTe$_2$/bilayer-In$_2$Se$_3$/Fe$_3$GeTe$_2$ heterostructure~\cite{yan2022giant}. Under the impetus of these studies, vdW-MFTJs have undergone explosive development~\cite{cao2022multiferroic,chen2022giant,zhang2023multilevel,zhu2023large,dong2023voltage,yang2023full,guo2024tunable,cui2024magnetic}.
However, alongside the flourishing development of vdW-MFTJs, two challenges have emerged that need to be overcome.
Firstly, the polarization direction of ferroelectric/ferromagnetic relies entirely on an externally applied electric/magnetic field in MFTJs, resulting in increased energy consumption. 
Secondly, the currently reported vdW-MFTJs are typically formed by combining ferroelectric and ferromagnetic materials to achieve multiferroicity, undoubtedly introducing more material interfaces and thereby increasing the complexity of experimental preparation.
Very recently, the emergence of sliding ferroelectric materials represented by the $h$-BN bilayer~\cite{li2017binary,wu2021two, FEBN} and the development of antiferromagnetic tunnel junctions~\cite{chen2023octupole,dong2022tunneling,qin2023room} have catered to the former challenge. The polarization direction flip of sliding ferroelectric materials only needs to overcome the weak van der Waals force and the antiferromagnetic tunnel junctions do not require a magnetic pinning layer, and smaller electric and magnetic fields are required.
Fortunately, bilayer VS$_2$ combines intrinsic multiferroicity, interlayer antiferromagnetism, and sliding ferroelectricity all in one~\cite{liu2020magnetoelectric}, making it an ideal candidate material for designing van der Waals intrinsic multiferroic tunnel junctions.

In this work, we theoretically design bilayer VS$_2$-based van der Waals intrinsic multiferroic tunnel junctions and investigate the electrode-dependent electronic transport properties by using first-principles computational methods. 
Here, we select three types of asymmetric electrode pairs, i.e., Ag/Au, 1T-MoS$_2$/2H-PtTe$_2$, and Fe$_3$GaTe$_2$/Fe$_3$GeTe$_2$.
Our calculation results reveal the emergence of non-volatile multiple states with giant TMR and TER, fostered by different types of electrodes, with the optimal electrode option being the van der Waals magnetic Fe$_3$GaTe$_2$/Fe$_3$GeTe$_2$ electrode pair.
Excitingly, we also observe perfect spin filtering, negative differential resistance effects, and an RA significantly less than 1 $\Omega$·$\mu$m$^{2}$ (The recording density of 200 Gbit/in$^{2}$ requires RA to be less than 1 $\Omega$·$\mu$m$^{2}$) within these intrinsic MFTJs.
Furthermore, we demonstrate bias- and stress-tunable electron transport properties, with TMR (TER) maximally increased to 34000\% (380\%) for Fe$_3$GaTe$_2$-VS$_2$-Fe$_3$GeTe$_2$ MFTJ and RA reduced to 0.015 $\Omega$·$\mu$m$^{2}$ for Ag-VS$_2$-Au MFTJ.
Additionally, the insertion of a monolayer graphene not only mitigates the impact of surface dangling bonds on Ag/Au electrodes but also further enhances TMR and TER by an order of magnitude.
The design concept of our intrinsic multiferroic tunnel junction holds promise for advancing the development of atomic-scale spintronics devices.  

\section{Computational Methods}
The structural relaxation, total energy, and electronic band structure calculations were conducted using density functional theory within the Vienna Ab initio Simulation Package (VASP)~\cite{VASP1996}. The project-augmented wave pseudopotentials method~\cite{PAW1994} and the general gradient approximation (GGA) in the Perdew-Burke-Ernzerhof (PBE) were adopted~\cite{GGA1992}. 
A cutoff energy of 500 eV and van der Waals correction using the DFT-D3 method were employed~\cite{DFTD3}. A $9\times9\times1$ Monkhorst–Pack $k$-grid~\cite{1976Special} was utilized to discretize the Brillouin zone of all VS$_2$-based heterojunction systems. 
In geometry optimization, the convergence criteria for electron energy and force are $10^{-5}$ eV and 0.01 eV/{\AA}, respectively.
The PBE+$U$ (on-site Coulomb interaction $U$$_{eff}$=3 eV for V atom) was used to treat localized $d$ orbitals. Note that the $U$ value for other material systems with $d$ orbitals, used as electrodes in this study, was not considered. Ferroelectric polarization is assessed employing the Berry phase method~\cite{king1993theory}.
The spin-polarized quantum transport properties are calculated within the framework of non-equilibrium Green's functions~\cite{NEGF2001} combined with density functional theory using the Nanodcal software~\cite{Nanodcal2001}. Also, the GGA with PBE function is employed for the electronic exchange-correlation function in the electron transport calculations.
In electronic self-consistent calculations, the cutoff energy was set to 80 Hartree, with a convergence criterion for the Hamiltonian matrix of $10^{-5}$ eV,  and the Fermi function temperature was set to 300 K.
A $k$-point grid of $100\times100\times1$ was employed to calculate current and electronic transmission coefficients. Biaxial strain is achieved by directly altering the in-plane lattice constants of the system.

The spin-polarized current $I_\sigma$ and conductance $G_\sigma$ are computed utilizing the Landauer-B\"uttiker formula \cite{CalcuI11992,book1995}:
\begin{align}
	I_\sigma &= \dfrac{e}{h} \int T_\sigma(E)[f_\text{L}(E) - f_\text{R}(E)] \mathrm{d}E,
\end{align}

\begin{align}
	G_\sigma &= \dfrac{e^{2}}{h}T_\sigma
\end{align}
Here, $\sigma$ denotes the spin index ($\uparrow,\downarrow$), $e$ is the electron charge, $h$ represents Planck's constant, $T_\sigma(E)$ stands for the spin-resolved transmission coefficient, and $f_\text{L(R)}(E)$ is the Fermi-Dirac distribution function of the left (right) lead.
The spin injection efficiency (SIE) is described by the following formula:
\begin{align}
	\rm SIE &=\left|\dfrac{I_\uparrow - I_\downarrow}{I_\uparrow + I_\downarrow}\right|.
\end{align}
The TMR at equilibrium can be defined as:~\cite{TMR1C}:
\begin{align}
	\rm TMR &=\frac{G_\text{PC}-G_\text{APC}}{G_\text{APC}}=\frac{T_\text{PC}-T_\text{APC}}{T_\text{APC}},
\end{align}
at bias voltage $V$, 
\begin{align}
	 \text{TMR}_{(V)} &=\frac{I_\text{PC}-I_\text{APC}}{I_\text{AP}},
\end{align}
where $T_\text{PC/APC}$ and $I_\text{PC/APC}$ represent the total transmission coefficient at the Fermi level and the currents under a bias voltage $V$ across the junctions in parallel configuration (PC) and antiparallel configuration (APC) magnetic states, respectively.
Another important physical quantity, TER, can be defined by the following equation~\cite{TERCalcu2016, FTJ0}:
\begin{align}
	\rm TER &=\frac{|G_\uparrow - G_\downarrow|}{\text{min}(G_\uparrow, G_\downarrow)}=\frac{|T_\uparrow - T_\downarrow|}{\text{min}(T_\uparrow, T_\downarrow)},
\end{align}
at bias voltage $V$, 
\begin{align}
	 \text{TER}_{(V)} &=\frac{|I_\uparrow - I_\downarrow|}{\text{min}(I_\uparrow, I_\downarrow)},
\end{align}
where $T_{\uparrow/\downarrow}$ and $I_{\uparrow/\downarrow}$ represent the total transmission coefficient at the Fermi level and currents under a bias voltage $V$, which can be obtained by reversing the direction of the ferroelectric polarization of the barrier layer.

The resistance-area (RA) product at equilibrium can be calculated from transmission by definition~\cite{Su2020NL}: 
\begin{align}
	\rm RA &=\frac{A}{G}=\frac{A}{{T_{(F)}} G_0},
\end{align}
at bias voltage $V$~\cite{yadav2022feal}, 
\begin{align}
	 \text{RA}_{(V)} &=\frac{V A}{I},
\end{align}
where $A$ is the unit cell area, $T_{(F)}$ is the calculated transmission at the Fermi level, and $G_0$=$e^2$/$h$ is the spin-conductance quantum,
$I$ is the current at bias voltage $V$.
\section{Results and discussion}
\subsection*{A. The design of vdW intrinsic MFTJs}
\begin{figure}[htb!]
	\centering
	\includegraphics[width=8.5cm,angle=0]{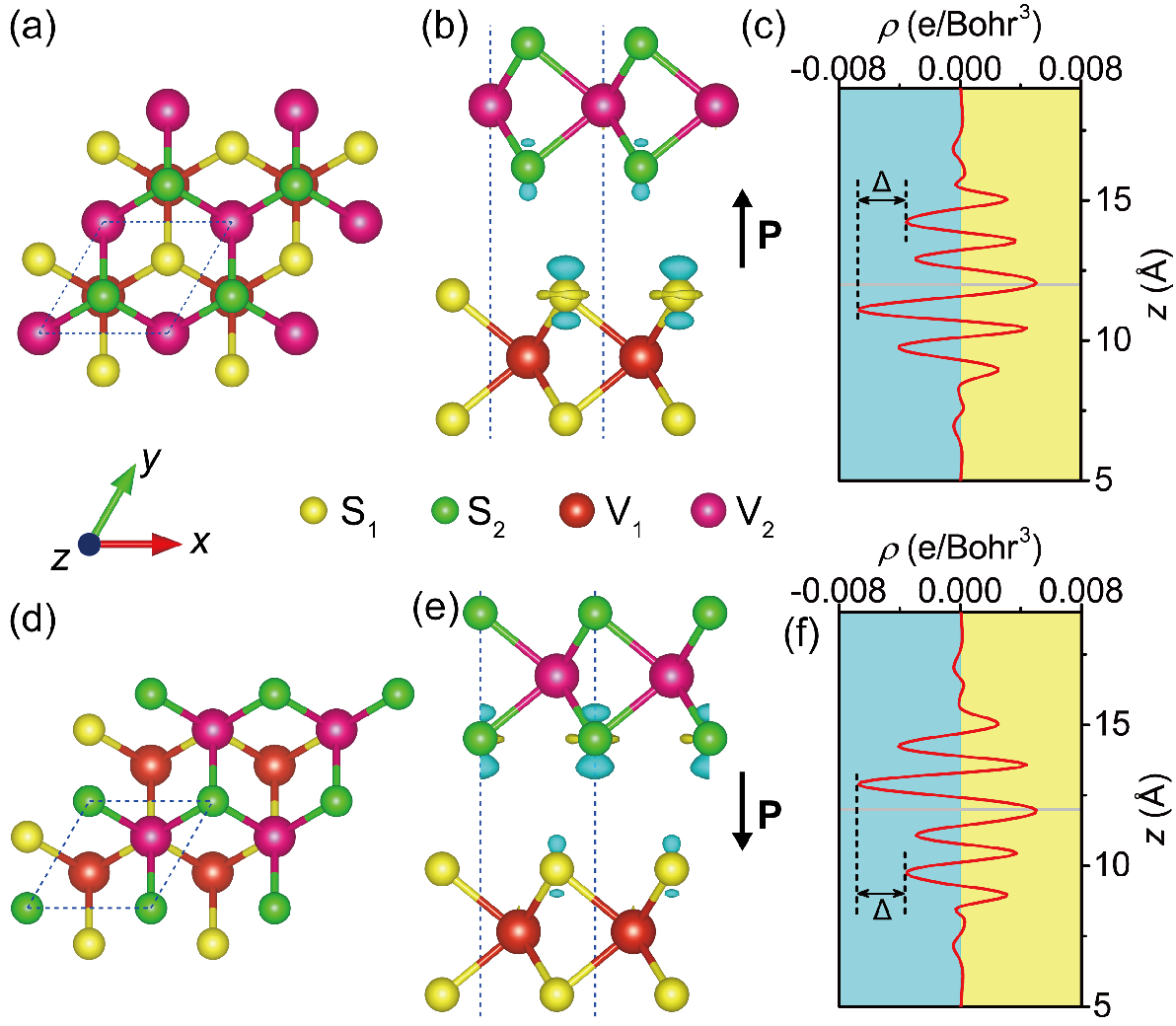}
	\caption{(a), (d) Top and (b), (e) side views of bilayer VS$_2$ crystal structures and the difference charge density for opposite out-of-plane ferroelectric polarization. The plane-averaged difference charge density along the $z$ direction of bilayer VS$_2$ with different ferroelectric polarization directions. The isosurface value is set to be 0.00025 $e$/Bohr$^3$. Yellow and blue colors denote charge accumulation and depletion.}
	\label{Fig1}
\end{figure}

The previous study~\cite{liu2020magnetoelectric} has indicated that the 3$R$-type (ground-state stacking of bilayer VS$_2$) stacking bilayer VS$_2$ exhibits intrinsic multiferroicity, characterized by interlayer antiferromagnetism and spontaneous out-of-plane ferroelectric polarization.
Moreover, polarization direction reversal can be achieved through interlayer sliding. Therefore, bilayer multiferroic VS$_2$ emerges as a candidate material for intrinsic MFTJs. The crystal structures of bilayer VS$_2$ with two opposite ferroelectric polarization directions are depicted in Fig.~\ref{Fig1}(a)-(b) and (d)-(e). 
To further confirm the multiferroicity of bilayer VS$_2$, 
we calculate that the total energy difference between its ferromagnetism and antiferromagnetism is 1.372 meV and its ferroelectric polarization value is determined to be $2.17\times10^{-3}$ C/m$^{2}$ using the Berry phase method~\cite{king1993theory}, which is consistent with the previous study~\cite{liu2020magnetoelectric}.
Fig.~\ref{Fig1}(b), (e) also include the calculated differential charge density, revealing a distinct difference in charge distribution between the accumulation and depletion regions of the upper and lower layers of VS$_2$. This inequivalence in charge results in a net charge transfer between the two layers, and the direction of the transfer reverses with changes in the stacking configuration, giving rise to opposing vertical polarization, which is further corroborated by the marked feature $\Delta$ in the plane-averaged differential charge density along the $z$-direction as depicted in the Fig~\ref{Fig1}(c), (f).

\begin{figure}[htb!]
	\centering
	\includegraphics[width=8.5cm,angle=0]{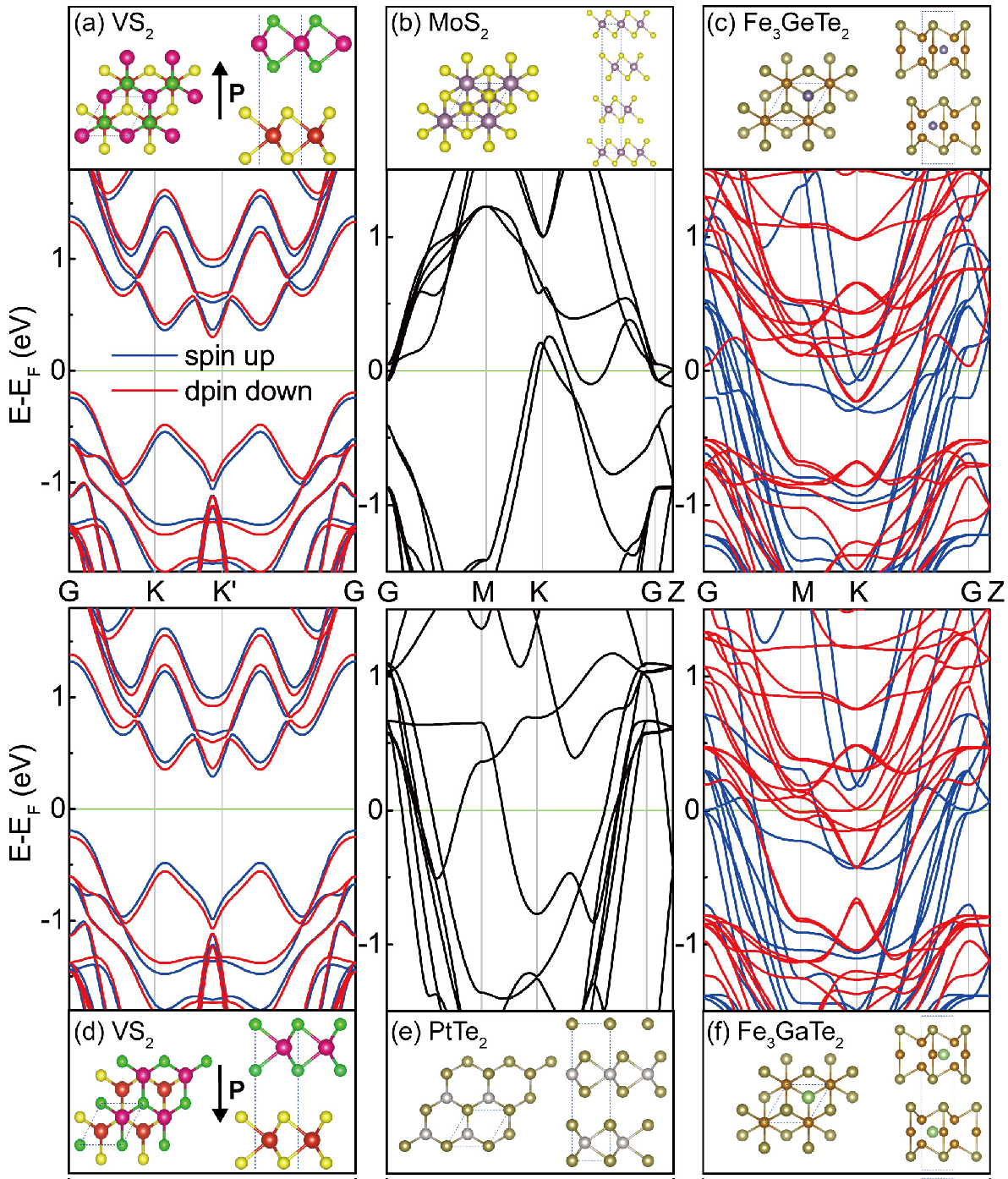}
	\caption{The Crystal structures and corresponding electronic band structures of (a) and (d) bilayer VS$_2$ with different ferroelectric polarizations, (b)/(e) bulk 1T-MoS$_2$/2H-PtTe$_2$, (c)/(f) bulk Fe$_3$GeTe$_2$/Fe$_3$GaTe$_2$.}
	\label{Fig2}
\end{figure}

\begin{figure*}[htp!]
	\centering
	\includegraphics[width=12cm,angle=0]{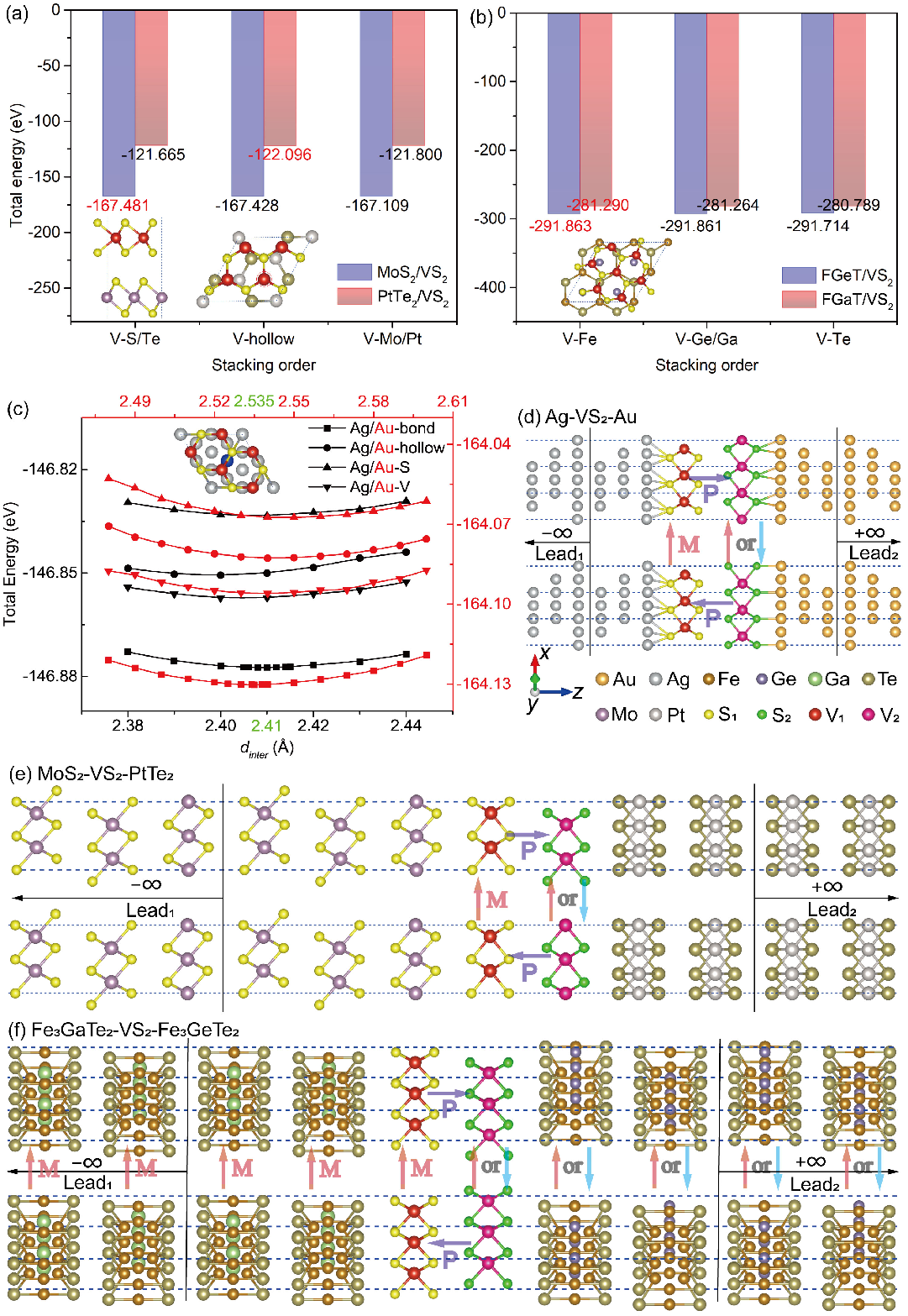}
	\caption{(a), (b) The total energy of 1T-MoS$_2$/2H-PtTe$_2$, Fe$_3$GaTe$_2$/Fe$_3$GeTe$_2$ interfaces $vs$ the various stacking orders. (c) The total energy of Ag/Au-VS$_2$ heterostructures with four stacking orders $vs$ the various interlayer distance $d$$_{inter}$. The inset represents the optimal stacking configuration. (d)-(f) Schematic structural diagrams of an intrinsic MFTJs device with opposite ferroelectric polarization directions composed of three types of asymmetric electrodes. (d) non-vdW pure metal Ag/Au, Ag-VS$_2$-Au MFTJ (e) vdW metallic 1T-MoS$_2$/2H-PtTe$_2$, 1T-MoS$_2$-VS$_2$-2H-PtTe$_2$ MFTJ (f) vdW ferromagnetic metallic Fe$_3$GaTe$_2$/Fe$_3$GeTe$_2$, Fe$_3$GaTe$_2$-VS$_2$-Fe$_3$GeTe$_2$ MFTJ. The left and right electrodes extend to $\mp\infty$. These MFTJs are periodic in the $xy$ plane and the current flows in the $z$ direction.}
	\label{Fig3}
\end{figure*}

\begin{table*}[htp!]
	\centering
	\renewcommand{\arraystretch}{1.8}
	\caption{Calculated spin-resolved electron transmission $T_{\uparrow}$ and $T_{\downarrow}$, TMR, TER, SIE, and RA at the equilibrium state for MFTJs with three different types of electrodes.}
	\label{table1}
 \resizebox{\linewidth}{!}{
\begin{tabular}{c c c c c c c c c c c c c c c}
\hline\hline
\multicolumn{2}{c}{\raisebox{-0.9em}{MFTJs}} & \multicolumn{2}{c}{\raisebox{-0.9em}{Polarization}} & \multicolumn{5}{c}{PC state (M${\uparrow\uparrow}$)} & \multicolumn{5}{c}{APC state (M${\uparrow\downarrow}$)} & \multirow{2}{*}{\large$\substack{\text{TMR}\\}$} \\ 
\cline{5-14}
\multicolumn{2}{c}{\raisebox{0.5em}{\ }} & \multicolumn{2}{c}{\raisebox{0.5em}{and Ratio}} & $T_{\uparrow}$ & $T_{\downarrow}$ & $T_\text{tot}=T_{\uparrow}+T_{\downarrow}$ & SIE & RA &  $T_{\uparrow}$ & $T_{\downarrow}$ & $T_\text{tot}=T_{\uparrow}+T_{\downarrow}$ & SIE & RA &  \\
\hline
\multicolumn{2}{c}{Ag-VS$_2$-Au} & \multicolumn{2}{c}{$\mathrm{P} \rightarrow$} & 0.0251 & 0.2191 & 0.2442 & 0.80 & 0.0278 & 0.1850 & 0.0658 & 0.2507 & 0.48 & 0.0256 & $-3 \%$ \\
\cline{3-15}
\multicolumn{2}{c}{\ } & \multicolumn{2}{c}{$\mathrm{P} \leftarrow$} & 0.0246 & 0.2398 & 0.2644 & 0.81 & 0.0270 & 0.2043 & 0.0783 & 0.2826 & 0.45 & 0.0240 & $-6 \%$ \\
\cline{3-15}
\multicolumn{2}{c}{\ } & \multicolumn{2}{c}{TER}& \multicolumn{5}{c}{8\%} & \multicolumn{5}{c}{7\%} &  \\
\hline
\multicolumn{2}{c}{MoS$_2$-VS$_2$-PtTe$_2$} & \multicolumn{2}{c}{$\mathrm{P} \rightarrow$} & 0.0025 & 0.0055 & 0.0080 & 0.38 & 1.0994 & 0.0091 & 0.0025 & 0.0116 & 0.56 & 0.7614 & $31 \%$ \\
\cline{3-15}
\multicolumn{2}{c}{\ } & \multicolumn{2}{c}{$\mathrm{P} \leftarrow$} & 0.0023 & 0.0068 & 0.0092 & 0.49 & 0.9638 & 0.0027 & 0.0018 & 0.0046 & 0.20 & 1.9254 & $100 \%$ \\
\cline{3-15}
\multicolumn{2}{c}{\ } & \multicolumn{2}{c}{TER}& \multicolumn{5}{c}{14\%} & \multicolumn{5}{c}{153\%} &  \\
\hline
\multicolumn{2}{c}{Fe$_3$GaTe$_2$-VS$_2$-Fe$_3$GeTe$_2$} & \multicolumn{2}{c}{$\mathrm{P} \rightarrow$} & 0.0324 & $2.36 \times 10^{-10}$ & $\sim0.0324$ & $\sim$1.00 & 0.4472 & $4.5 \times 10^{-4}$ & $8.99 \times 10^{-6}$ & $4.56 \times 10^{-4}$ & 0.96 & 31.7640 & $7003 \%$ \\
\cline{3-15}
\multicolumn{2}{c}{\ } & \multicolumn{2}{c}{$\mathrm{P} \leftarrow$} &  0.0298 & $1.43 \times 10^{-8}$ & $\sim0.0298$ & $\sim$1.00 & 0.4866 & 0.0002 & 0.0001 & 0.0003 & 0.46 & 53.7661 & $10949 \%$ \\
\cline{3-15}
\multicolumn{2}{c}{\ } & \multicolumn{2}{c}{TER}& \multicolumn{5}{c}{9\%} & \multicolumn{5}{c}{69\%} &  \\
\hline\hline
\end{tabular}
}
\end{table*}

Upon confirming the multiferroicity in bilayer VS$_2$, we can construct intrinsic multiferroic tunnel junctions (MFTJs) based on VS$_2$. Firstly, we calculate the electronic band structure of the bilayer VS$_2$ with opposite polarization directions and the corresponding crystal structure is shown in Fig.~\ref{Fig2}(a), (d).
It is observed that, although the ground state of bilayer VS$_2$ is antiferromagnetic, the band structure exhibits a slight band splitting due to the presence of ferroelectric polarization~\cite{liu2020magnetoelectric}. Additionally, it is found to be an indirect bandgap semiconductor. Therefore, bilayer VS$_2$ can serve both as a ferroelectric barrier layer and a magnetic layer, making it suitable for the central scattering region of MFTJs devices.
Subsequently, the choice of electrode materials needs to be determined. To meet the requirements of an asymmetric structure on both sides of the central scattering region in the ferroelectric tunnel junction, we select three types of asymmetric electrode pairs: non-vdW pure metal electrodes (Ag/Au), vdW non-magnetic electrodes (1T-MoS$_2$/2H-PtTe$_2$), and vdW magnetic electrodes (Fe$_3$GaTe$_2$/Fe$_3$GeTe$_2$). In addition to the pure metal electrode pair Ag/Au, Fig.~\ref{Fig2}(b) and (e), and (c) and (f) illustrate the bulk crystal structures and corresponding electronic band structures of these electrodes, respectively. One can observe that the Fermi level intersects with the entire band structure, indicating metallic properties for these materials and making them suitable for use as electrodes.

After determining the electrodes and the multiferroic barrier layer, we can now construct three types of intrinsic MFTJ devices, namely, Ag/VS$_2$/Au, 1T-MoS$_2$/VS$_2$/2H-PtTe$_2$, and Fe$_3$GaTe$_2$/VS$_2$/Fe$_3$GeTe$_2$ MFTJs.
The pure metal Ag/Au(111) has a hexagonal lattice with an in-plane lattice constant of 2.889/2.884 {\AA} in the unit cell.
Consistent with the previous studies, the optimized in-plane lattice constants of monolayer 1T-MoS$_2$, 2H-PtTe$_2$, Fe$_3$GaTe$_2$, Fe$_3$GeTe$_2$, and VS$_2$ are 3.143 {\AA}\cite{ekspong2020theoretical}, 3.895 {\AA}~\cite{kempt2020two}, 4.026 {\AA}~\cite{lee2023electronic}, 4.020 {\AA}~\cite{FGTlatticecon}, and 3.157 {\AA}~\cite{zhuang2016stability} respectively. 
Therefore, Ag/Au(111), 1T-MoS$_2$, 2H-PtTe$_2$, and Fe$_3$GaTe$_2$/Fe$_3$GeTe$_2$ have in-plane minimum supercell matches with VS$_2$ of $2\times2$/$2\times2$(Ag/Au)@$\sqrt{3}\times\sqrt{3}$(VS$_2$), $2\times2$@$2\times2$, $\sqrt{3}\times\sqrt{3}$@$2\times2$, and $2\times2$/$2\times2$@$\sqrt{7}\times\sqrt{7}$, respectively.
Considering the primary focus on the intrinsic multiferroicity of bilayer VS$_2$, the lattice constant of VS$_2$ is taken as the in-plane lattice constant for all MFTJs as a whole.
In this case, the maximum in-plane mismatch ratios obtained for the three types of MFTJs are 5.2\%, 6.4\%, and 3.75\%, respectively.
These mismatch ratios are unlikely to occur in realistic experiments due to the weak vdW interactions between layers in the van der Waals systems~\cite{MTJ2,dong2023voltage}. However, they must be considered in theoretical simulations due to the periodic boundary conditions.
In the following section, we also investigate the influence of in-plane biaxial strain on the transport properties of these MFTJs.
Next, we need to determine the optimal stacking configurations at various interfaces of these MFTJs.
There are six types of heterojunction interfaces in these MFTJs, i.e., Ag/Au-VS$_2$, MoS$_2$/PtTe$_2$-VS$_2$, Fe$_3$GaTe$_2$/Fe$_3$GeTe$_2$-VS$_2$. In Ag/Au-VS$_2$, four high-symmetry stacking configurations (Ag/Au-bone/hollow/S/V) exist. Fig.~\ref{Fig3}(c) illustrates the evolution of stacking configurations with the interlayer distance ($d_{inter}$), revealing that the lowest-energy stacking configuration and interlayer distance are designated as Ag/Au-bone and 2.41/2.535 {\AA}, respectively, with the corresponding optimal structures presented in the inset.
For the MoS$_2$/PtTe$_2$-VS$_2$ heterojunction, we consider five stacking configurations (V-S/Te, V-hollow, and V-Mo/Pt), as illustrated in Fig.~\ref{Fig3}(a). The energetically favored stacking configuration is identified as V-S/hollow. Similarly, as shown in Fig.~\ref{Fig3}(b), one can determine that the optimal stacking arrangement for Fe$_3$GaTe$_2$/Fe$_3$GeTe$_2$-VS$_2$ is V-Fe/Fe.
Finally, based on the stacking sequences of the interfaces established above, we can construct three different intrinsic MFTJs, labeled as Ag-VS$_2$-Au, MoS$_2$-VS$_2$-PtTe$_2$, and Fe$_3$GaTe$_2$-VS$_2$-Fe$_3$GeTe$_2$, as depicted in Fig.~\ref{Fig3}(d)-(f).
Note that a complete atomic relaxation was performed on the central scattering region with a vacuum layer thickness of 30 {\AA}  of these MFTJ devices.

\subsection*{B. Significant TMR/TER and biaxial strain effects at equilibrium}
By the physical mechanism of the magnetic/ferroelectric tunnel junction, flipping the magnetization direction of one layer of VS$_2$ can lead to two opposing magnetic configurations (PC/APC), while sliding the ferroelectric bilayer VS$_2$ results in two polarization directions. This implies the potential induction of a quadruple resistive state in the VS$_2$-based MFTJs.
In this section, we first investigate the TMR and TER effects of these three types of intrinsic MFTJs at equilibrium.
As presented in \autoref{table1}, the transport properties of MFTJs with different electrodes show considerable differences.
Distinguished from the other two, the Ag-VS$_2$-Au MFTJ displays negative TMR, being -3\%/-6\% in the ferroelectric P$\rightarrow/$P$\leftarrow$ state. For the MoS$_2$-VS$_2$-PtTe$_2$ MFTJ, its TMR is larger than that of Ag-VS$_2$-Au MFTJs but much smaller than that of Fe$_3$GaTe$_2$-VS$_2$-Fe$_3$GeTe$_2$.
Among the three, the highest TMR (7003\%/10949\% in P$\rightarrow$/P$\leftarrow$ state) is achieved in the Fe$_3$GaTe$_2$-VS$_2$-Fe$_3$GeTe$_2$ MFTJ and a perfect spin polarization (SIE=$\sim$1) in the PC state.
In addition to the TMR effect, the TER effect is crucial for assessing the transport performance of MFTJ devices.
From \autoref{table1}, it is evident that the maximum TER of 153\% occurs in the MoS$_2$-VS$_2$-PtTe$_2$ MFTJ at APC state, but it is not significantly larger than the other two. We also calculate the resistance-area (RA) product of these MFTJs. 
It is worth mentioning that the calculated RA products of all four resistance states of Ag-VS$_2$-Au MFTJ are less than 0.03 $\Omega$·$\mu$m$^{2}$, which is an ideal characteristic of MFTJ for device applications. 
Additionally, the RA in the Fe$_3$GaTe$_2$-VS$_2$-Fe$_3$GeTe$_2$ MFTJ is also considerable, being less than 1 $\Omega$·$\mu$m$^{2}$. 
This contrasts sharply with the previously calculated RA products for perovskite-oxide MFTJs of around several $k\Omega$·$\mu$m$^{2}$~\cite{borisov2015spin,velev2009magnetic}.
Therefore, overall, the MFTJ composed of vdW magnetic electrodes excels over the other two counterparts due to its giant TMR, larger TER, perfect spin filtering, and RA product less than 1 $\Omega$·$\mu$m$^{2}$.

\begin{figure}[htp!]
	\centering		
	\includegraphics[width=8.0cm,angle=0]{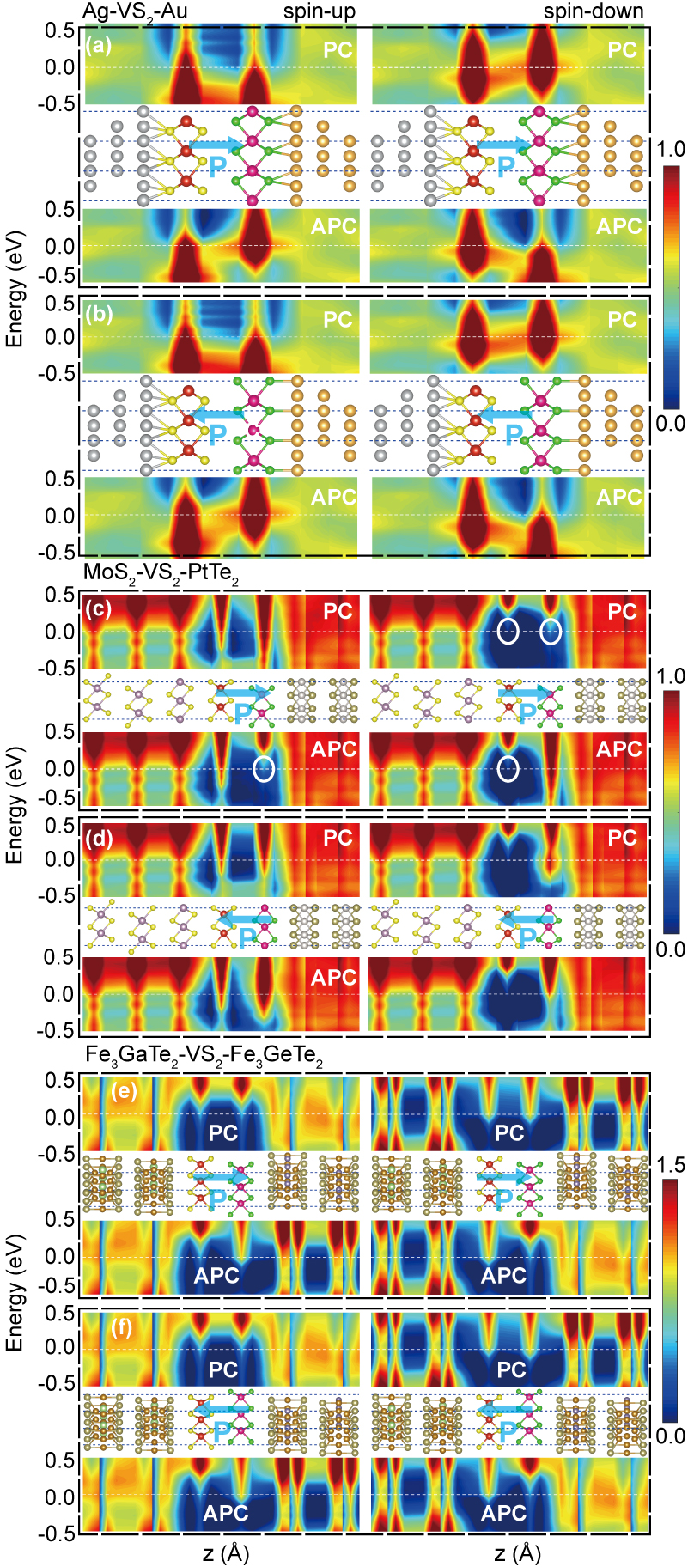}
	\caption{Spin-resolved PDOS and the corresponding crystal structure of the central scattering region along the transport direction $z$-axis for intrinsic MFTJs with three different types of electrodes in the equilibrium state. White dashed lines mark the Fermi level. The white ovals in panels (c) and (d) mark the minority-state electron density.}
	\label{Fig4}
\end{figure}

\begin{figure*}[htp!]
	\centering	
	\includegraphics[width=18cm,angle=0]{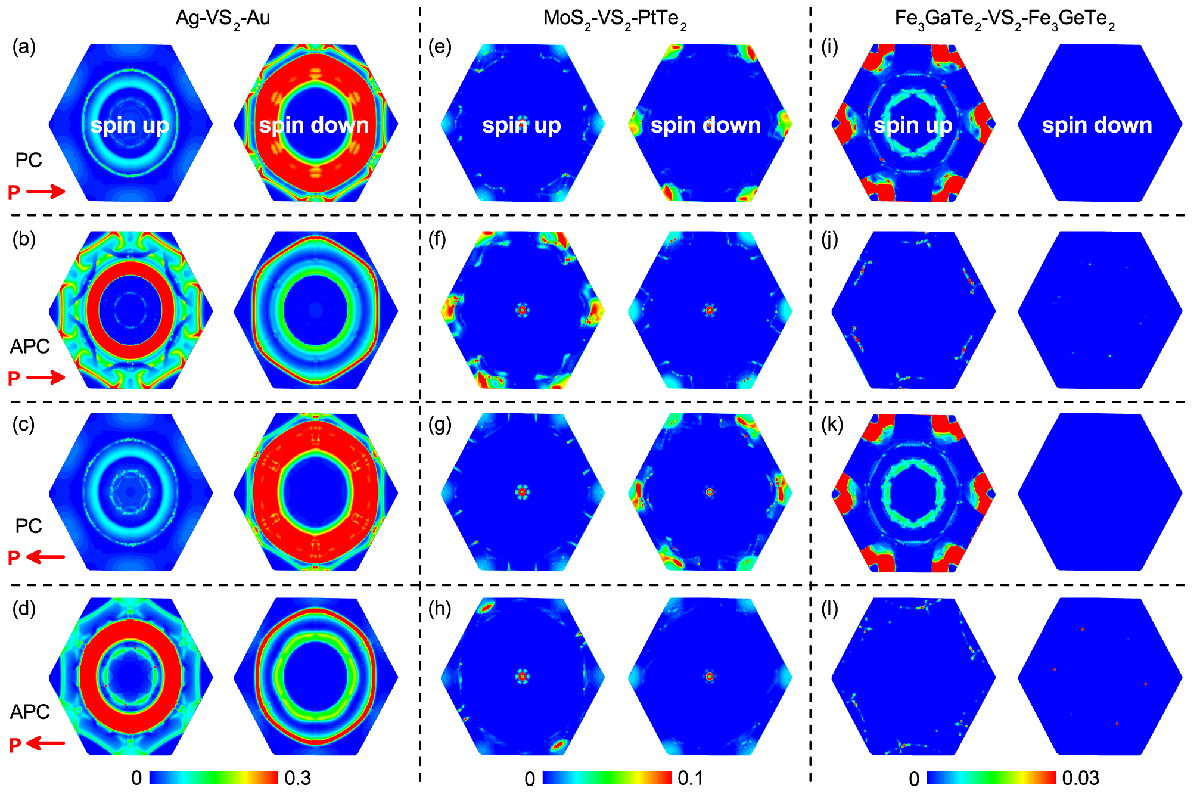}
	\caption{The $k_\Arrowvert$-resolved transmission coefficients across the three different VS$_2$-based MFTJs in the 2D Brillouin zone for {P$\rightarrow$/P$\leftarrow$} and {PC(M$\uparrow\uparrow$)/APC(M$\uparrow\downarrow$)} states at the Fermi level. (a)-(d) Ag-VS$_2$-Au MFTJ (e)-(h) 1T-MoS$_2$-VS$_2$-2H-PtTe$_2$ MFTJ (i)-(l) Fe$_3$GaTe$_2$-VS$_2$-Fe$_3$GeTe$_2$ MFTJ.}
	\label{Fig5}
\end{figure*}

To provide a more detailed insight into the multiple nonvolatile resistance states in these MFTJs, we analyze the Partial Density of States (PDOS) of the central scattering region in the ($E$, $z$) plane. Here, $E$ represents the Fermi energy, and $z$ signifies the vertical distance along the transport direction. Figure~\ref{Fig4} illustrates the spin-resolved PDOS of the central scattering region in three types of MFTJs, considering opposite ferroelectric polarization and magnetic alignment directions. For the Ag-VS$_2$-Au MFTJ, the presence of dangling bonds on the Ag/Au surface forming chemical bonds with VS$_2$ leads to a modification in the electronic structure of VS$_2$. As shown in Fig.~\ref{Fig4}(a) and (b), there is a significant overlap of wave functions between the bilayer VS$_2$ at the Fermi level, indicating that this MFTJ does not operate through electron tunneling mechanism, resulting in very low TMR and TER.
In Section. D, the insertion of a monolayer graphene between VS$_2$ and the metal electrodes Ag/Au not only effectively shields the influence of the dangling bonds but also preserves the multiferroic properties of VS$_2$.
In the other two MFTJs, the little electron density in the deep blue region between the bilayer VS$_2$ at the Fermi level indicates that electron transport occurs through a tunneling mechanism [see Fig.~\ref{Fig4}(c)-(f)].
Clearly, a typical TMR effect is evident in the PDOS diagram.
We use MoS$_2$-VS$_2$-PtTe$_2$ MFTJ as a representative to reveal the TMR effect.
Note that here we consider the spin-down DOS of VS$_2$ at parallel configuration (PC) state as the minority state, as indicated by white circles in Fig.~\ref{Fig4}(c) and (d).
As shown in Fig.~\ref{Fig4}(c), for the antiparallel configuration (APC) state, electrons with spin-up (spin-down) flow from the left VS$_2$ with majority (minority) states at the Fermi level, then flow out from the right VS$_2$ with minority (majority) states. The corresponding relationship between the opposite electron state densities in the bilayer VS$_2$ implies the high-resistance state. For the PC state, spin-up electrons are in the majority state in both the upper and lower layers of VS$_2$, resulting in low-resistance transport. On the contrary, the VS$_2$ in the upper and lower layers of spin-down electrons are in minority states near the Fermi level, which hinders electron transport. The above analysis can be directly applied to the Fe$_3$GaTe$_2$-VS$_2$-Fe$_3$GeTe$_2$ MFTJ.

\begin{figure*}[htp!]
	\centering	
	\includegraphics[width=18cm,angle=0]{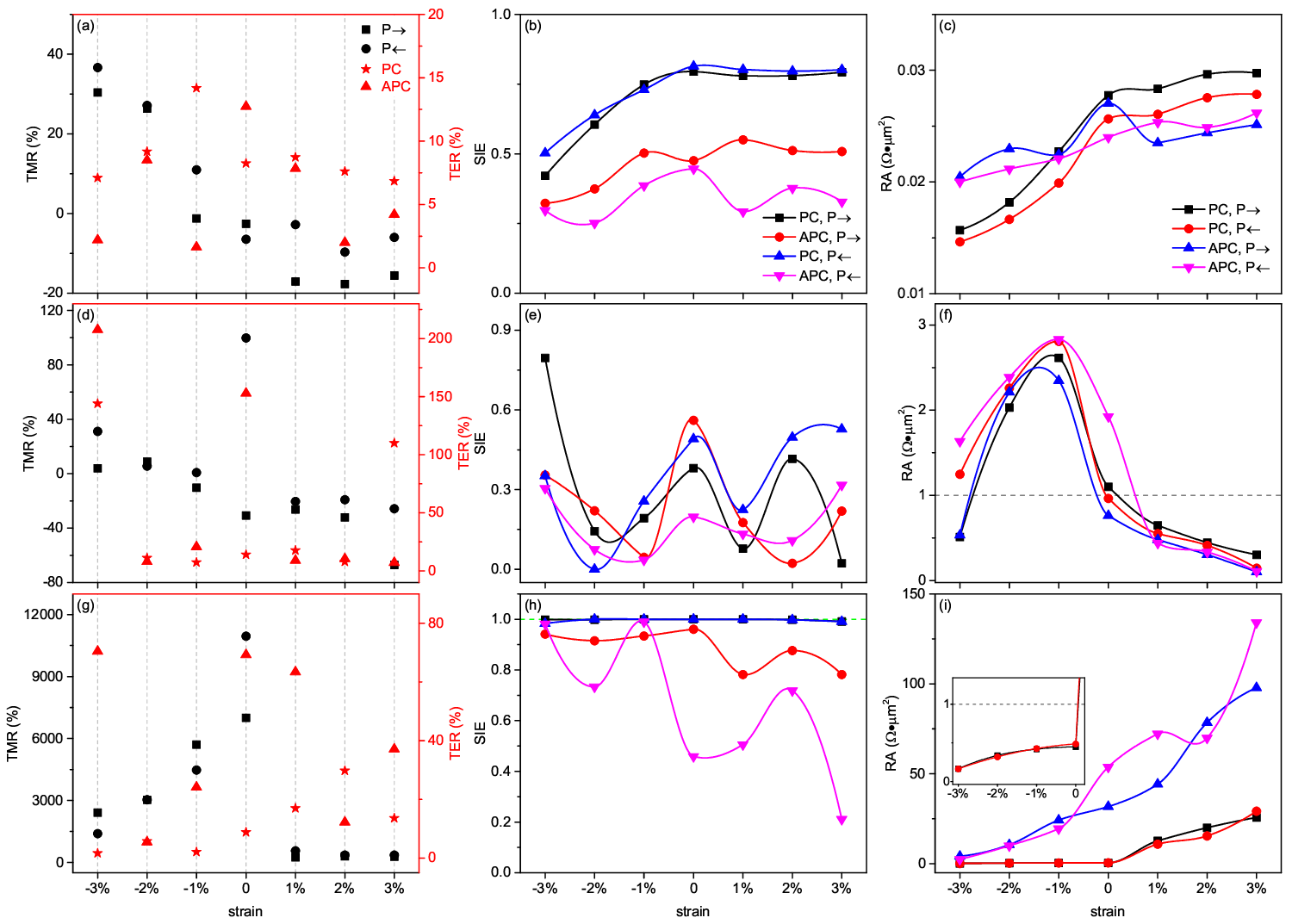}
	\caption{The TMR [(a), (d), and (g)], SIE [(b), (e) and (h)], and RA [(c), (f), and (i)] as functions of in-plane biaxial strain. Strain effect in the equilibrium state of (a)-(c) Ag-VS$_2$-Au MFTJ; (d)-(f) 1T-MoS$_2$-VS$_2$-2H-PtTe$_2$ MFTJ; (g)-(i) Fe$_3$GaTe$_2$-VS$_2$-Fe$_3$GeTe$_2$ MFTJ.}
	\label{Fig6}
\end{figure*}

\begin{figure*}[htp!]
	\centering	
	\includegraphics[width=18cm,angle=0]{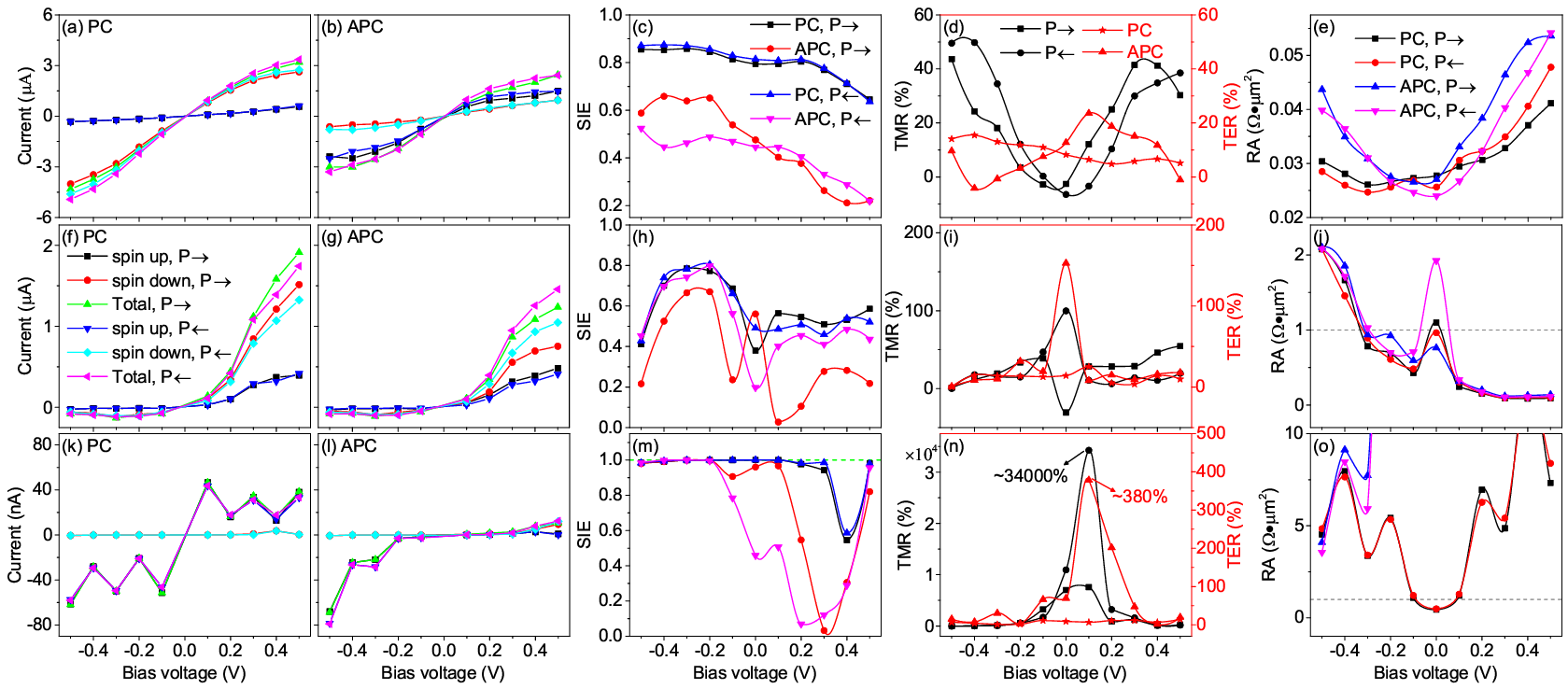}
	\caption{The variation of the current [(a), (b), (f), (g), (k) and (l)], spin injection efficiency (SIE) [(c), (h) and (m)], TMR and TER ratios [(d), (i) and (n)], and RA [(e), (j) and (o)] as a function of the bias voltages for three different MFTJs. Transport properties of (a)-(e) Ag-VS$_2$-Au MFTJ; (f)-(j) 1T-MoS$_2$-VS$_2$-2H-PtTe$_2$ MFTJ; (k)-(o) Fe$_3$GaTe$_2$-VS$_2$-Fe$_3$GeTe$_2$ MFTJ.}
	\label{Fig7}
\end{figure*}

To further elucidate how ferroelectric polarization and magnetization alignment influence electron transmission, we calculate the $k_\Arrowvert$-resolved transmission coefficients of these MFTJs at the Fermi level within the 2D Brillouin zone (2D-BZ), as depicted in Fig.~\ref{Fig5}, which are perpendicular to the transport direction ($z$-axis). 
As shown in Fig.~\ref{Fig5}(a)-(d), overall, for the Ag-VS$_2$-Au MFTJ, the difference in the "hot spots" between PC/P$\rightarrow$ and APC/P$\leftarrow$ is small, corresponding to negligible TMR and TER  (both less than 10\%).
The maximum TER (153\%) is achieved in the MoS$_2$-VS$_2$-PtTe$_2$ MFTJ at the APC state, as clearly revealed in the 2D-BZ electron transmission spectra at the Fermi level. 
Specifically, by comparing Fig.~\ref{Fig5}(f) and (h), one can observe that in the P$\rightarrow$ state, the "hot spots" for spin-up are more concentrated near the high-symmetry points $K$ and $K'$ than in the P$\leftarrow$ state, while the "hot spots" for spin-down are slightly larger in P$\rightarrow$ than in P$\leftarrow$. This leads to a significant TER, as indicated by Equation (6).
For the Fe$_3$GaTe$_2$-VS$_2$-Fe$_3$GeTe$_2$ MFTJ, a significantly giant TMR and perfect spin filtering effect can be reflected from the electron transmission coefficients at the Fermi level, with the maximum TMR reaching 10949\% in the ferroelectric polarization P$\leftarrow$ state. 
As depicted in Fig.~\ref{Fig5}(k) and (l), a substantial number of "hot spots" emerge in the spin-up channel of the PC state, with almost no "hot spots" in the spin-down and APC configurations. This suggests the potential presence of significant TMR and high spin polarization rates.
Therefore, the distribution of transmission coefficients in the 2D-BZ offers further evidence supporting the pres
ence of a giant TMR/TER ratio and a perfect spin-filtering effect in these MFTJs in the equilibrium state.

Previous studies have suggested that applying in-plane biaxial strain is an effective measure to enhance the transport performance of vdW MFTJs~\cite{yan2022giant,dong2023voltage}. Considering the mismatch at the interfaces of our MFTJs, we systematically investigate the influence of in-plane biaxial strain in the range of -3\% to 3\% with a 1\% interval on these MFTJs in the equilibrium state, and the results are presented in Fig.~\ref{Fig6}. It can be observed that biaxial strain significantly modulates TMR, TER, SIE, and RA. For Ag-VS$_2$-Au MFTJ, as shown in Fig.~\ref{Fig6}(a), regardless of P$\rightarrow$ or P$\leftarrow$ states, TMR increases with increasing compressive strain, and the TMR value changes from negative to positive. Conversely, tensile strain has a minor effect, maintaining negative values throughout.
In contrast to TMR, overall TER decreases with increasing strain. From Fig.~\ref{Fig6}(b), it can be observed that in the PC state, the SIE of P$\rightarrow$/P$\leftarrow$ remains around 80\% unaffected by tensile strain, but decreases with increasing compressive strain.
Figure~\ref{Fig6}(c) indicates that the RA of the MFTJ in the quadruple resistance state generally decreases during the transition from high tensile strain to compressive strain.
For MoS$_2$-VS$_2$-PtTe$_2$ MFTJ, as displayed in Fig.~\ref{Fig6}(d), only at a strain of -3\%, the TMR becomes positive, while under other strains, it remains negative. Meanwhile, the TER reaches its maximum value at this compressive strain (-3\%), approximately 200\%.
Furthermore, the SIE exhibits oscillatory behavior, peaking at 80\% under a strain of -3\% [see Fig.~\ref{Fig6}(e)]. Interestingly, as shown in Fig.~\ref{Fig6}(f), the RA for all four resistance states peaks at over 3 $\Omega$·$\mu$m$^{2}$ under a strain of -1\%. Above or below this strain value, the RA rapidly decreases to a minimum of approximately 0.1 $\Omega$·$\mu$m$^{2}$. 
Exciting stain effects are observed in the Fe$_3$GaTe$_2$-VS$_2$-Fe$_3$GeTe$_2$ MFTJ. As displayed in Fig.~\ref{Fig6}(g), both TMR values for the two ferroelectric polarization states (P$\rightarrow$/P$\leftarrow$) increase with increasing compressive strain, while tensile stress remains unchanged. The strain has little effect on TER, with small numerical oscillations. 
Fig.~\ref{Fig6}(h) reflects that the SIE under the PC state remains constant at 100\% within the studied strain range, indicating the robustness of the perfect spin filtering effect against strain. Like the Ag-VS$_2$-Au MFTJ, RA gradually decreases to 0.16 $\Omega$·$\mu$m$^{2}$ during the transition from maximum tensile to compressive strain.
Therefore, the aforementioned results indicate that strain serves as an effective approach for modulating the transport properties of MFTJs.

\subsection*{C. Voltage-tunable transport properties in nonequilibrium state}
Next, as shown in Fig.~\ref{Fig7}, we calculate the bias voltage-dependent (ranging from -0.5 V to 0.5 V) spin polarization current, spin injection efficiency (SIE), TMR ratio, TER ratio, and RA of these tunnel junctions in the PC(M$\uparrow\uparrow$)/APC(M$\uparrow\downarrow$) and P$\rightarrow$/P$\leftarrow$ states.
Note that the bias voltage, denoted as $V{_b}$, is set by applying the chemical potential on the left (right) electrode as $+V_b/2$ ($-V_b/2$).
For the Ag-VS$_2$-Au MFTJ, as depicted in Fig.~\ref{Fig7}(a), it is evident that in the PC state, regardless of the ferroelectric polarization orientation, the current monotonically increases with increasing bias voltage, and the total current is mostly contributed by the spin-down channel [with the SIE greater than 0.8 in Fig.~\ref{Fig7}(c)], indicating a significant spin polarization rate. Under the APC state, the current exhibits a monotonic increase, but the rate of increase is slower and smaller compared to the PC state [see Fig.~\ref{Fig7}(b)].
As depicted in Fig.~\ref{Fig7}(d), one can observe that the TMR increases with bias voltage in both P$\rightarrow$ and P$\leftarrow$ states, and TMR becomes positive for bias voltages exceeding 0.2 V. The maximum TMR is less than 50\%, and the bias voltage has a relatively minor impact on TER. Excitingly, as shown in Fig.~\ref{Fig7}(e), despite the increase in the resistance-area product (RA) with increasing bias voltage across the four resistance states, the maximum value remains below 0.06 $\Omega$·$\mu$m$^{2}$. Such a small value is beneficial for device performance.

Let us turn to the MoS$_2$-VS$_2$-PtTe$_2$ MFTJ.
As shown in Figs~\ref{Fig7}(f) and (g), the current-voltage (IV) characteristics of the PC and APC states resemble a diode-like behavior, where the diode conducts under positive bias voltage and exhibits reverse bias leakage current.
At the equilibrium state, the SIE of this MFTJ is only around 0.5 (refer to \autoref{table1}). However, when the bias voltage is set to -0.2 V, it can be increased to above 0.8 as shown in Fig~\ref{Fig7}(h). 
It is evident that bias voltage has an adverse effect on both TMR and TER, with the maximum values occurring at zero bias voltage [see Fig~\ref{Fig7}(i)]. 
Interestingly, as shown in Fig~\ref{Fig7}(j), the RA across the four resistance states shows synchronized evolution under the different bias voltages. That is, it monotonically increases with positive bias voltage, reaching a minimum value of 0.11 $\Omega$·$\mu$m$^{2}$, while under negative bias voltage, it initially decreases and then increases, indicating the beneficial effect of positive bias voltage on enhancing the performance of this tunnel junction.

\begin{figure*}[htp!]
	\centering	
	\includegraphics[width=18cm,angle=0]{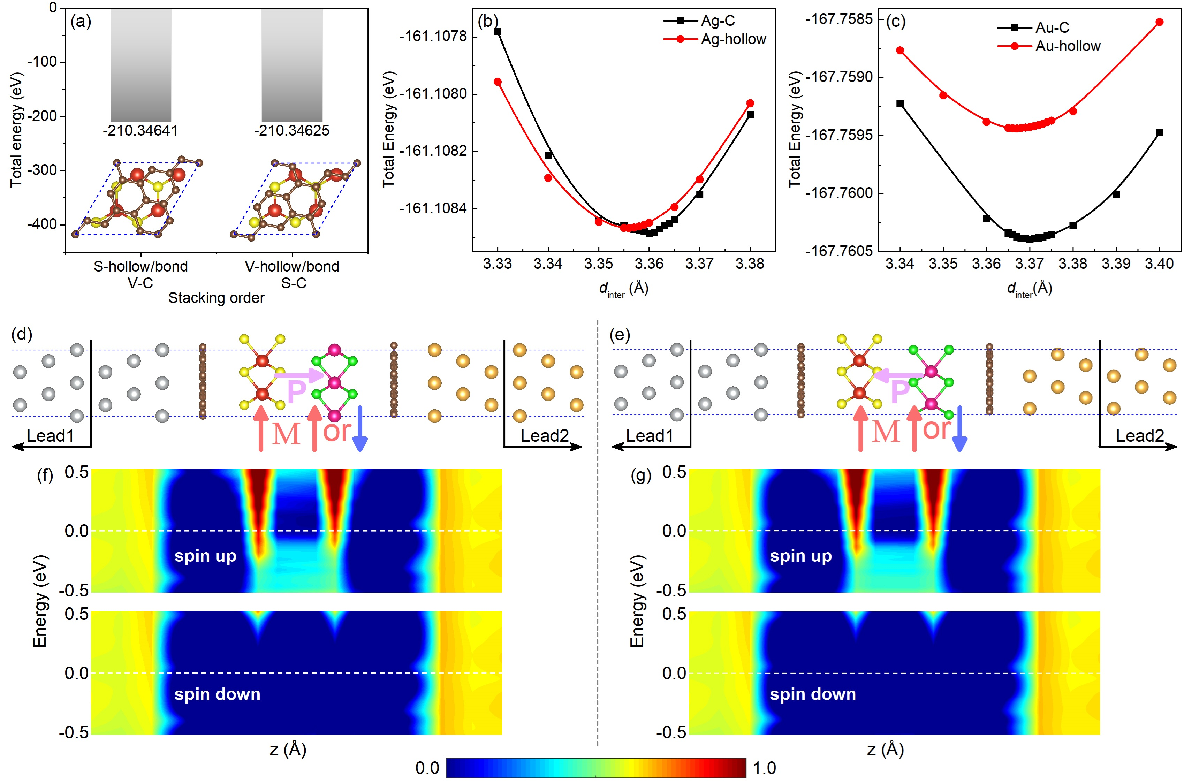}
	\caption{(a) The total energy of Graphene(Gra)-VS$_2$ interface versus the various stacking orders. (b) and (c) The total energy of Ag/Au-Gra heterostructures with two stacking orders $vs$ the various interlayer distance $d$$_{inter}$. [(d), (e)] and [(f), (g)] Schematic diagrams and spin-polarized PDOS in the central scattering region along the transport direction $z$-axis at PC(M$\uparrow\uparrow$) state in the equilibrium state of Ag-Gra-VS$_2$-Gra-Au MFTJs devices with P$\rightarrow$ and P$\leftarrow$ states.}
	\label{Fig8}
\end{figure*}

\begin{table*}[htp!]
	\centering
	\renewcommand{\arraystretch}{1.8}
	\caption{Calculated spin-dependent electron transmission $T_{\uparrow}$ and $T_{\downarrow}$, TMR, TER, SIE, and RA at the equilibrium state for Ag-Gra-VS$_2$-Gra-Au MFTJ.}
	\label{table2}
 \resizebox{\linewidth}{!}{
\begin{tabular}{c c c c c c c c c c c c c c c}
\hline\hline
\multicolumn{2}{c}{\raisebox{-0.9em}{MFTJ}} & \multicolumn{2}{c}{\raisebox{-0.9em}{Polarization}} & \multicolumn{5}{c}{PC state (M${\uparrow\uparrow}$)} & \multicolumn{5}{c}{APC state (M${\uparrow\downarrow}$)} & \multirow{2}{*}{\large$\substack{\text{TMR}\\}$} \\ 
\cline{5-14}
\multicolumn{2}{c}{\raisebox{0.5em}{\ }} & \multicolumn{2}{c}{\raisebox{0.5em}{and Ratio}} & $T_{\uparrow}$ & $T_{\downarrow}$ & $T_\text{tot}=T_{\uparrow}+T_{\downarrow}$ & SIE & RA &  $T_{\uparrow}$ & $T_{\downarrow}$ & $T_\text{tot}=T_{\uparrow}+T_{\downarrow}$ & SIE & RA &  \\
\hline
\multicolumn{2}{c}{Ag-Gra-VS$_2$-Gra-Au} & \multicolumn{2}{c}{$\mathrm{P} \rightarrow$} & $6.83\times10^{-4}$ & $6.30\times10^{-3}$ & $6.98\times10^{-3}$ & 0.80 & 1.2951 & 0.0165 & 0.0181 & 0.0346 & 0.05 & 0.2610 & -80\% \\
\cline{3-15}
\multicolumn{2}{c}{\ } & \multicolumn{2}{c}{$\mathrm{P} \leftarrow$} & $9.63\times10^{-4}$ & 0.0121 & 0.0131 & 0.85 & 0.6898 & 0.0189 & 0.0201 & 0.0390 & 0.03 & 0.2317 & -66\% \\
\cline{3-15}
\multicolumn{2}{c}{\ } & \multicolumn{2}{c}{TER}& \multicolumn{5}{c}{88\%} & \multicolumn{5}{c}{13\%} &  \\
\hline\hline
\end{tabular}
}
\end{table*}

\begin{figure*}[htp!]
	\centering	
	\includegraphics[width=17cm,angle=0]{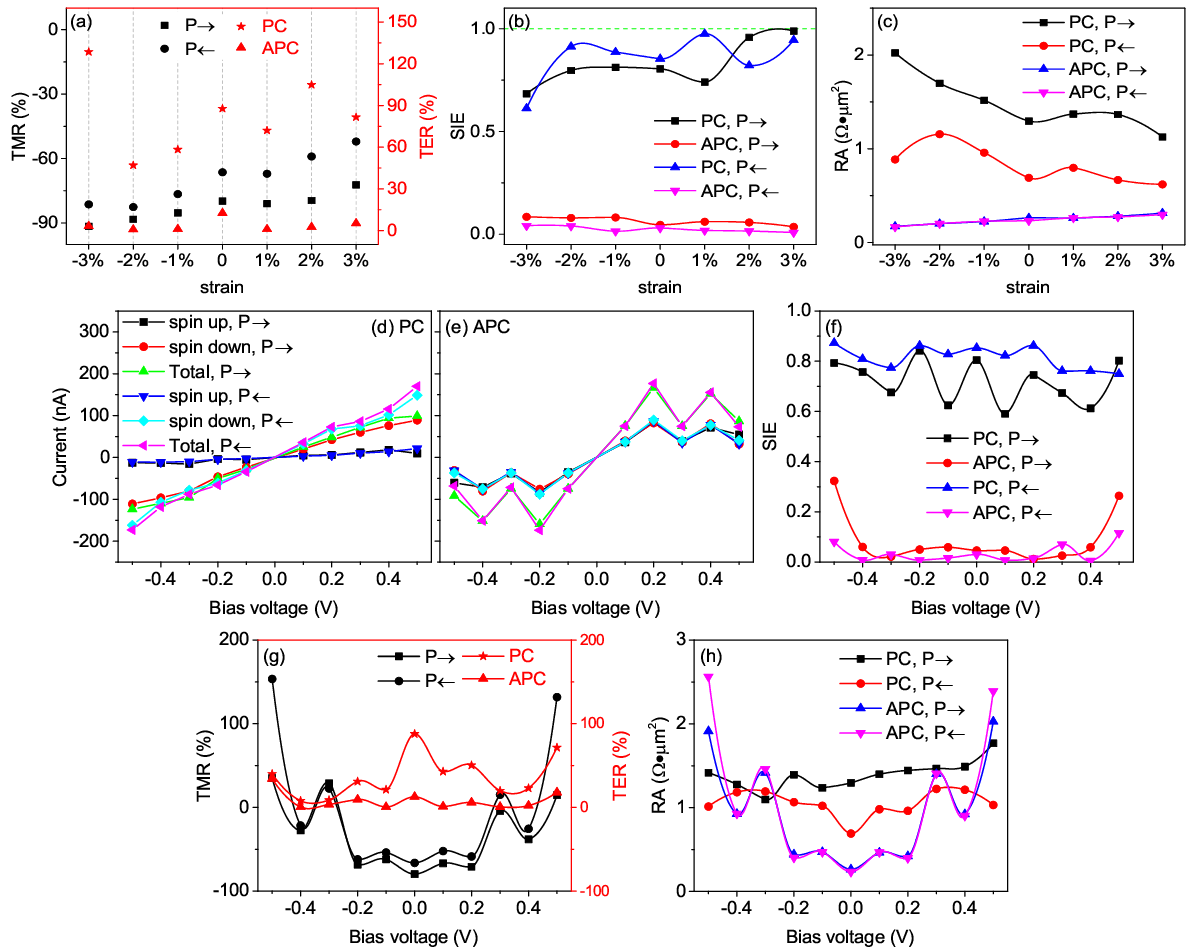}
	\caption{The TMR (a), SIE (b), and RA (c) as functions of in-plane biaxial strain for Ag-Gra-VS$_2$-Gra-Au MFTJ in the equilibrium state. The variation of the current [(d), (e)], spin injection efficiency (SIE) (f), TMR and TER ratios (g), and RA (h) as a function of the bias voltages for Ag-Gra-VS$_2$-Gra-Au MFTJ in the non-equilibrium state.}
	\label{Fig9} 
\end{figure*}

The most significant bias voltage effect occurs in the Fe$_3$GaTe$_2$-VS$_2$-Fe$_3$GeTe$_2$ MFTJ.
In the PC state, whether it is P$\rightarrow$ or P$\leftarrow$ state, the current exhibits oscillatory behavior with increasing bias voltage, and the total current is contributed by the spin up current, implying negative differential resistance and perfect spin filtering effect [see Fig.~\ref{Fig7}(k)]. 
Compared to the PC state, as shown in Fig.~\ref{Fig7}(l), the current in the APC state remains nearly zero within the bias voltage range of -0.2 V to 0.2 V, indicating the presence of a giant magnetoresistance effect.
As shown in Fig.~\ref{Fig7}(m), one can find that the SIE of the P$\rightarrow$/P$\leftarrow$ under the PC state exhibits robustness against bias voltage within the range of -0.5 V to 0.3 V and maintains a value close to 1. This implies that the perfect spin filtering effect is not constrained by bias voltage, indicating strong controllability of spin current. 
More significantly, the bias voltage can markedly improve TMR and TER, as shown in Fig.~\ref{Fig7}(n). In both ferroelectric polarization P$\rightarrow$/P$\leftarrow$ states, the TMR decreases with increasing negative bias voltage. However, it reaches its maximum (7600\%/34000\%) when the positive bias voltage increases to 0.1 V, after which it decreases rapidly. Under the PC state, the TER remains almost unaffected by the bias voltage, maintaining a constant value. However, under the APC state, the TER follows a similar trend to the evolution of TMR under both ferroelectric polarizations. The maximum TER occurs at 0.1 V and can reach up to 380\%. In addition, the application of bias voltage in Fig.~\ref{Fig7}(o) is obviously detrimental to RA, but fortunately, it is less than 1 $\Omega$·$\mu$m$^{2}$ in the range of -0.1 V to 0.1 V.

Based on the above discussion, we can obtain that the Fe$_3$GaTe$_2$-VS$_2$-Fe$_3$GeTe$_2$ MFTJ, under bias voltage, exhibits the optimal performance among tunnel junction devices due to its combination of maximum TMR/TER, perfect spin filtering, and negative differential resistance effects. Therefore, in comparison to the other two types of electrodes, we consider the van der Waals magnetic electrode to be the best choice. Note that considering lattice matching, we selected Fe$_3$GaTe$_2$/Fe$_3$GeTe$_2$ materials as magnetic electrodes, which inherently possess a high spin polarization. For MTJs/MFTJs, magnetic materials with high spin polarization will yield large TMR. Therefore, choosing magnetic materials with half-metal properties will result in even better transport properties. 
\subsection*{D. Graphene intercalation effect in Ag-Gra-VS$_2$-Gra-Au MFTJ}
For the Ag-VS$_2$-Au MFTJ, in accordance with the aforementioned discussions, the strong hybridization between Ag/Au and adjacent VS$_2$ weakens the charge redistribution ferroelectricity between the bilayer VS$_2$ and alters the intrinsic electronic structure, ultimately resulting in its minimal TMR and TER.
To preserve the intrinsic multiferroicity of bilayer VS$_2$, we introduce monolayer graphene (Gra) interlayers between Ag/Au and VS$_2$, referring to previous studies~\cite{dong2023voltage,yang2022giant}. The geometric structure of the new MFTJ device is denoted as Ag-Gra-VS$_2$-Gra-Au as shown in Figs.~\ref{Fig8}(d) and (e).
In this new MFTJ, there are three different interfaces: Ag-Gra, Au-Gra, and VS$_2$-Gra.
To determine the stacking configurations of these interfaces, we calculate the evolution of their total energy with stacking orders and interlayer distance, as shown in Figs.~\ref{Fig8}(a)-(c). From the inserted structure diagram in Fig.~\ref{Fig8}(a), one can observe that the VS$_2$-Gra heterostructure exhibits two sets of three equivalent stacking sequences: one comprising S-hollow/bond and V-C, and the other consisting of V-hollow/bond and S-C.
The calculation results of the total energy indicate that the first combination (V-C) is the optimal stacking configuration for VS$_2$-Gra interface.
For the Ag/Au-Gra interface, we calculate the evolution of the total energy with the interlayer distance for two stacking orders, as shown in Figs.~\ref{Fig8}(b) and (c). It can be observed that the optimal stacking order and interlayer distance are Ag/Au-C and 3.36/3.37 {\AA}, respectively.
The PDOS of the central scattering region in real space along the transport direction can clearly reflect the effect of graphene intercalation, as depicted in Figs.~\ref{Fig8}(f) and (g). Note that here we are only presenting the PDOS of the PC states under equilibrium conditions.
From the spin up PDOS, it can be observed that a broad blue-black region emerges between the metal electrodes Ag/Au and VS$_2$, indicating a clear isolation of the wave function coupling between the two.
Additionally, one can observe a blue-black region between the bilayer VS$_2$ at the Fermi level. Contrasting with the mentioned above PDOS of Ag-VS$_2$-Au MFTJ, this reveals that graphene intercalation preserves the intrinsic properties of the bilayer VS$_2$.

\autoref{table2} summarizes the transport properties of the Ag-Gra-VS$_2$-Gra-Au MFTJ MFTJ at the equilibrium state. 
Clearly, under the influence of graphene intercalation, the values of all four non-volatile resistive states have been elevated by an order of magnitude.
We also investigate the transport characteristics of this MFTJ under the influence of applied biaxial stress and bias voltage, as shown in Fig.~\ref{Fig9}. The TMR of both ferroelectric polarization states increases with increasing compressive stress (ignoring sign) and decreases with increasing tensile stress [see Fig.~\ref{Fig9}].
In general, apart from a 2\% tensile stress, all other strain conditions have a weakening effect on TER. As depicted in Fig.~\ref{Fig9}(b), tensile strain can elevate the SIE close to 1 for the PC state, while it has almost no effect on the APC state.
The smaller the RA of the MFTJ, the better its performance. It is evident that strain cannot significantly reduce RA compared to the case without strain [see Fig.~\ref{Fig9}(c)].
Figure~\ref{Fig9}(d) shows that under the PC state, the total current for both ferroelectric polarization arrangements P$\rightarrow$/P$\leftarrow$ approaches linear increase with increasing bias voltage, with the majority contributed by the spin-down current.
In the APC state, the current exhibits oscillatory behavior with bias voltage and surpasses the current under the PC state at certain biases, implying a negative TRM effect [refer to Fig.~\ref{Fig9}(e)].
As shown in Fig.~\ref{Fig9}(f), throughout the entire range of bias voltages, the SIE of the APC state is significantly smaller than that of the PC state.
Specifically, the bias voltage can increase the maximum TMR to about 150\%(-0.5 V), while the maximum TER occurs under non-biased conditions [see Fig.~\ref{Fig9}(g)]. For RA, as shown in Fig.~\ref{Fig9}(h), the RA of four resistive states increases under almost all bias voltages, indicating a detrimental effect on the performance of the MFTJ.

Therefore, our analysis demonstrates that graphene intercalation is an effective strategy to enhance the transport performance of MFTJs and preserve the intrinsic multiferroic properties of bilayer VS$_2$. This finding could pave the way for novel experimental approaches.

\section{Summary}
In summary, based on the first-principles density functional theory, we theoretically study the spin-dependent electronic transport properties of the bilayer VS$_2$-based vdW intrinsic MFTJs. 
Three different types of asymmetric electrode pairs are employed to investigate the influence on the transport properties of MFTJs.
We demonstrate that the four giant non-volatile resistance states and low resistance-area products (RA) in these MFTJs depend on electrode selection and the potential to manipulate these states by applying bias voltage and in-plane biaxial strain.
At the equilibrium state, the maximum achievable TMR (TER) due to electrode effects is 10949\% (153\%), with the minimum RA being 0.026 $\Omega$·$\mu$m$^{2}$. Strain effects enhance TER to 208\%, with RA decreasing to 0.016 $\Omega$·$\mu$m$^{2}$. Under non-equilibrium conditions, bias voltage further enlarges TMR (TER) to 34000\% (380\%).
Comparing the three types of electrodes, we reveal that the electrode composition consisting of Fe$_3$GaTe$_2$/Fe$_3$GeTe$_2$ offers the optimal choice for MFTJs due to the combination of giant TMR (TER), perfect spin filtering, and negative differential resistance effects.
Furthermore, we also prove that graphene intercalation not only effectively eliminates the disruption of the hanging bonds in the Ag/Au electrodes for bilayer VS$_2$ multiferroicity but also enhances the TMR (TER) by an order of magnitude.
This work provides new design concepts for van der Waals intrinsic multiferroic tunnel junctions and paves the way for next-generation low-energy spintronics memory devices.

\section{Acknowledgements}
This work was supported by the National Key Research and Development Program of China (No. 2022YFB3505301), the National Natural Science Foundation of China (No. 12304148), the Natural Science Basic Research Program of Shanxi (No. 202203021222219 and No. 202203021212393), and the Project funded by China Postdoctoral Science Foundation (No.2023M731452). Z.Y. thanks X. Shi (from HZWTECH) for his valuable discussions.

\bibliography{main}
\bibliographystyle{apsrev4-2}
\end{document}